\documentclass[twocolumn,showpacs,amsmath,amssymb,prl,superscriptaddress,floatfix]{revtex4-1}
\usepackage{graphicx}
\usepackage{bm}
\usepackage{bbold}
\usepackage{amsmath,amsfonts,amssymb}

\begin{document}

\title{Characterization of dynamical phase transitions in quantum jump trajectories beyond the properties of the stationary state}

\author{Igor Lesanovsky}
\affiliation{School of Physics and Astronomy, University of
Nottingham, Nottingham, NG7 2RD, UK}
\author{Merlijn van Horssen}
\affiliation{School of Mathematical Sciences, University of
Nottingham, Nottingham, NG7 2RD, UK}
\author{Madalin Guta}
\affiliation{School of Mathematical Sciences, University of
Nottingham, Nottingham, NG7 2RD, UK}
\author{Juan P. Garrahan}
\affiliation{School of Physics and Astronomy, University of
Nottingham, Nottingham, NG7 2RD, UK}


\date{\today}

\begin{abstract}
We describe how to characterize dynamical phase transitions in open quantum systems from a purely dynamical perspective, namely, through the statistical behavior of quantum jump trajectories. This approach goes beyond considering only properties of the steady state. While in small quantum systems dynamical transitions can only occur trivially at limiting values of the controlling parameters, in many-body systems they arise as collective phenomena and within this perspective they are reminiscent of thermodynamic phase transitions.  We illustrate this in open models of increasing complexity: a three-level system, a dissipative version of the quantum Ising model, and the micromaser. In these examples dynamical transitions are accompanied by clear changes in static behavior. This is however not always the case, and in general dynamical phase behavior needs to be uncovered by observables which are strictly dynamical, e.g. dynamical counting fields. We demonstrate this via the example of a class of models of dissipative quantum glasses, whose dynamics can vary widely despite having identical (and trivial) stationary states.
\end{abstract}

\maketitle
Recent experimental progress in quantum optics and cold atomic physics has stimulated a great interest in the study of open many-body quantum systems \cite{Diehl10,DallaTorre10,*Pichler10,Lee2012,*Foss-Feig12,Ates12,Kessler12,Marcos12,Nagy11}. Currently, much effort is dedicated to the understanding and classification of dynamical phases and transitions among them, for example in the Dicke Model \cite{Clive03,Baumann09,*Keeling10,Nagy11}, in lattice bosons subject to engineered dissipation \cite{Diehl10}, and in spin systems \cite{Lee2012,*Foss-Feig12,Ates12,Kessler12,Olmos12}. Typically, dynamical phase transitions are detected and analyzed through changes in static order parameters, such as superfluid density or spin polarization, calculated in the system's stationary state.

The aim of this work is to characterize dynamical phases exclusively through time-correlations within quantum jump trajectories, and not through static order parameters. We do so by building on an elegant connection between open quantum systems -- described by a Lindblad master equation -- and matrix product states \cite{Affleck&Kennedy&Lieb,*Fannes&Nachtergale&Werner} (MPS) put forward in \cite{Schon05,*Verstraete10,*Osborne&Eisert&Verstraete10}. We pursue the usual dual description of open system dynamics: On the one hand any individual realization of the dynamics is stochastic. In the case of an open quantum system this is represented by a stochastic wave function corresponding to a specific sequence of quantum jumps, with the whole ensemble of these stochastic trajectories being encoded in a MPS. On the other hand, the evolution of probabilities is deterministic and is derived from the evolution of the density matrix under the action of a quantum master operator (QMO). The dynamical phase structure of an open system is given by the low lying spectrum of the QMO \cite{Kessler12}. More specifically, dynamical phases are characterized by how the spectrum responds to changes in physical parameters \cite{Lee2012,*Foss-Feig12,Ates12,Kessler12} or counting fields \cite{Lecomte2007,*Garrahan2007,Garrahan10}. When this response is non-analytic we have the signature of a dynamical phase transition \cite{Lecomte2007,*Garrahan2007,Garrahan10}. This analytic structure is, however, also mirrored in the ensemble of trajectories. So in order to understand dynamical phase transitions we need to study both the spectrum of the QMO and the MPS that encodes the trajectories.

We illustrate our approach through a number of examples exhibiting a wide variety of dynamical features. To set the stage we first discuss three well-studied open systems that would appear to exhaust the range of possibilities: a ``blinking'' three-level system \cite{Plenio1998,Garrahan10}, a dissipative quantum Ising model with a dynamical first-order transition \cite{Lee2012,Ates12}, and the micromaser, which has dynamical transitions of both first and second order kind  \cite{Benson94,*Briegel&Englert&Sterpi&Walther,*Rempe&Walther90,*Englert02,Garrahan11}.  In all these cases transitions in dynamics are related to changes in static behavior.  We consider however a fourth class of problems, illustrated via a model of dissipative quantum glasses \cite{Olmos12}, whose dynamical behavior can change abruptly while their statics remain invariant throughout. This highlights the need for a dynamical approach like the one presented here for characterizing complex open systems where static order parameters are non-existent or difficult to identify.

The density matrix $\rho$ of the open quantum systems we consider evolves under the Master equation $\partial_t\rho=\mathcal{W}(\rho)$ with the QMO given by $\mathcal{W}(\bullet)\equiv\mathcal{H}(\bullet)+\mathcal{D}(\bullet)$. Here the super-operators $
\mathcal{H}(\bullet)=-i\left[H,\bullet\right]$ and $\mathcal{D}(\bullet)=\sum_{m=1}^N L_{m} \bullet L_{m}^\dagger-\frac{1}{2}\sum_{m=1}^N\left\{L_m^\dagger L_{m},\bullet\right\}$ govern the coherent and dissipative dynamics, respectively. They depend on the hamiltonian $H$ and the set $\{ L_m, m=1,\ldots,N \}$ of jump operators. The specific form of the latter depends on the coupling of the system to the bath. It is well established that an open system dynamics generates a MPS on the bath degrees of freedom \cite{Schon05}. We use this connection to link the analysis of the emission sequence of the bath quanta, i.e. the quantum jump trajectories, to the more familiar picture of static phases of the ground state of a one-dimensional spin system.
The prescription of Refs.\ \cite{Schon05} is most clear for the evolution of the density matrix $\rho$ over short but finite time intervals $\delta t$, represented for instance by the Kraus map
\begin{eqnarray}
T_{\delta t} (\rho) \equiv e^{\mathcal{H}\delta t} e^{\mathcal{D}\delta t} (\rho)= K_0 \rho K_0^\dagger + \sum_{m=1}^N K_m \rho K^\dagger _m ,
\label{eq:map}
\end{eqnarray}
with Kraus operators $K_0=e^{-i \delta t H}\, \sqrt{1-\delta t\sum_{k=1}^N L^\dagger_k L_k}$ and $K_m=e^{-i \delta t H}\,\sqrt{\delta t }L_m$. Such discrete (but non-unique) representations of dissipative evolutions  have been employed recently in experiment to simulate the dynamics of open many-body systems \cite{Bareiro11}.

\begin{figure}
\includegraphics[width=\columnwidth]{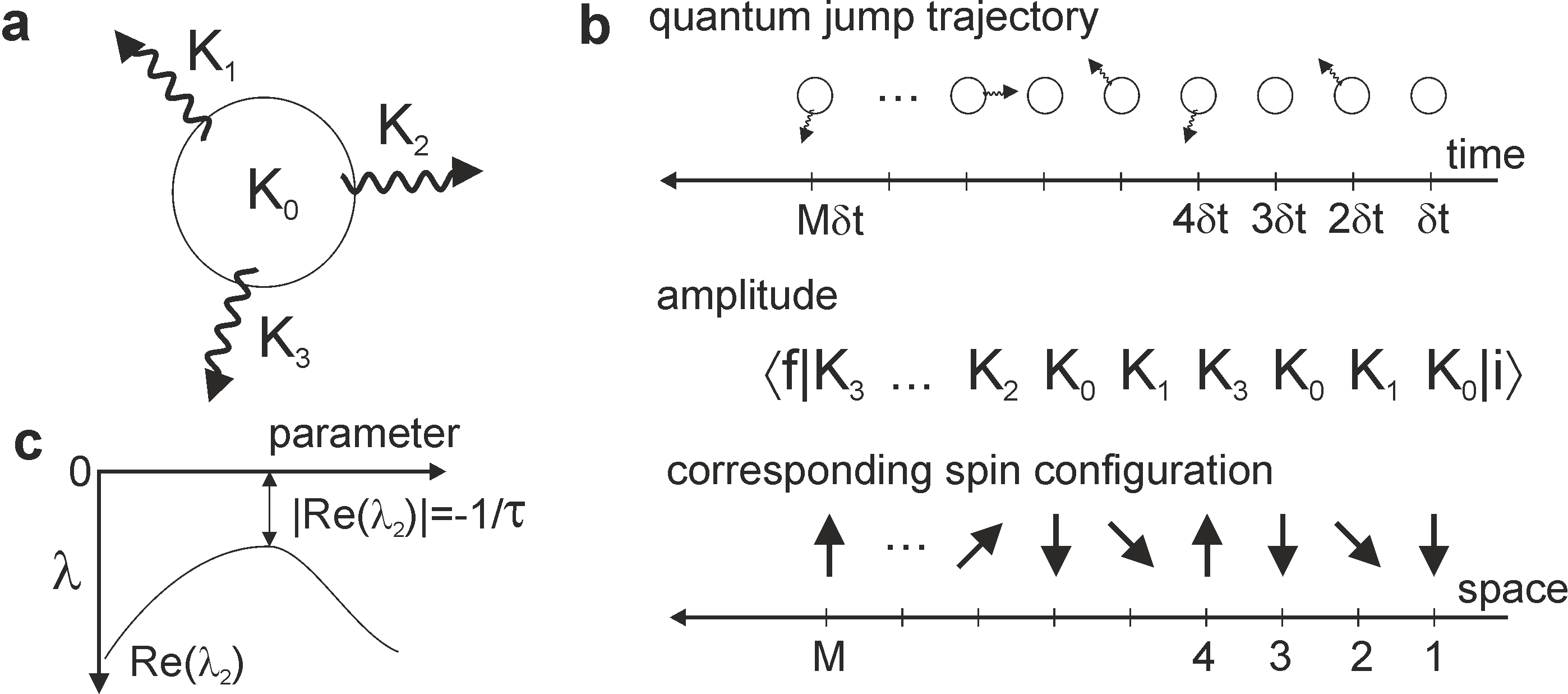}
\caption{\textbf{a:} Kraus operators defining open dynamics: $K_0$ is the non-unitary (no-jump) evolution, while $K_{m>0}$ refers to the occurrence of a quantum jump of kind $m$. \textbf{b:} The amplitude for a sequence of quantum jumps is obtained directly from the Kraus operators. This can be mapped to the state of a fictitious one-dimensional spin system, and all possible trajectories with their amplitudes can be gathered in an MPS interpreted as the quantum state of the spin chain. \textbf{c:} The two-time correlation time of quantum jumps (or correlation length of the spin chain) is the inverse of the spectral gap of the QMO $\mathcal{W}$; dynamical transitions occur when this gap closes as a function of an external parameter.}
\label{fig:time_v_space}
\end{figure}

Fig.\ \ref{fig:time_v_space}a sketches the action of the Kraus operators: $K_0$ corresponds to non-unitary (i.e.\ no-jump) evolution, while $K_{m>0}$ represents the effect of the quantum jump associated to $L_{m}$.  The evolution of the density matrix over macroscopic times is generated by multiple applications of the map (\ref{eq:map}). This produces quantum jump trajectories as shown in the upper panel of Fig. \ref{fig:time_v_space}b.  The probability for a certain trajectory $\{n_1,n_2,\dots ,n_M\}$ to occur after $M$ time steps is given by
$p_{n_1 n_2 \dots n_M}=
 \sum_{f} |\langle f \mid K_{n_M} \dots K_{n_2}K_{n_1} \mid i \rangle|^2$
, where $|i\rangle$ is the initial state of the system. The sum runs over a basis of final system states $\mid f \rangle$, and each term is the probability for connecting the initial and final states via a certain sequence of quantum jumps.

These probabilities are encoded in an MPS which can be thought of as being generated by letting the system interact sequentially with a chain of
$(N+1)$-dimensional spins initially prepared in a fixed pure state \cite{Schon05}. After $M$ steps the quantum state of the system \emph{and} the $M$ spins with which it has interacted is
$|\Psi\rangle = \sum_{f} |f\rangle \otimes | \psi(f)\rangle$,
where $|\psi(f)\rangle$ is the (unnormalized) MPS
\begin{equation}
 | \psi(f)\rangle=  \sum_{n_M,...,n_1=0}^N  \left<f\right| K_{n_M}\dots K_{n_1}\left|i\right> \left|n_1,..., n_M\right> ,
 \label{eq:MPS}
\end{equation}
with the sum running over all spin basis vectors.
We can think of $|\psi(f)\rangle$ as the ground state of a fictitious one-dimensional spin system with specific boundary conditions (see Fig.\ \ref{fig:time_v_space}b).
The state $|\Psi\rangle$ therefore encodes the whole {\em ensemble} of quantum trajectories: each basis state $\left|n_1,..., n_M\right>$ corresponds to a specific trajectory and its amplitude $\left<f\right| K_{n_M}\dots K_{n_1}\left|i\right>$ is related to the probability $p_{n_1 n_2 \dots n_M}$ of it occurring dynamically.
While this connection is only formal it illustrates that the study of dynamical phases of open systems is not different from that of the static properties of a one-dimensional spin system, regardless of the actual spatial dimension of the open problem. Dynamical phase transitions will then become visible in the time correlations of the quantum jumps which correspond to spatial correlations in the spins.

\begin{figure*}
\includegraphics[width=1.8\columnwidth]{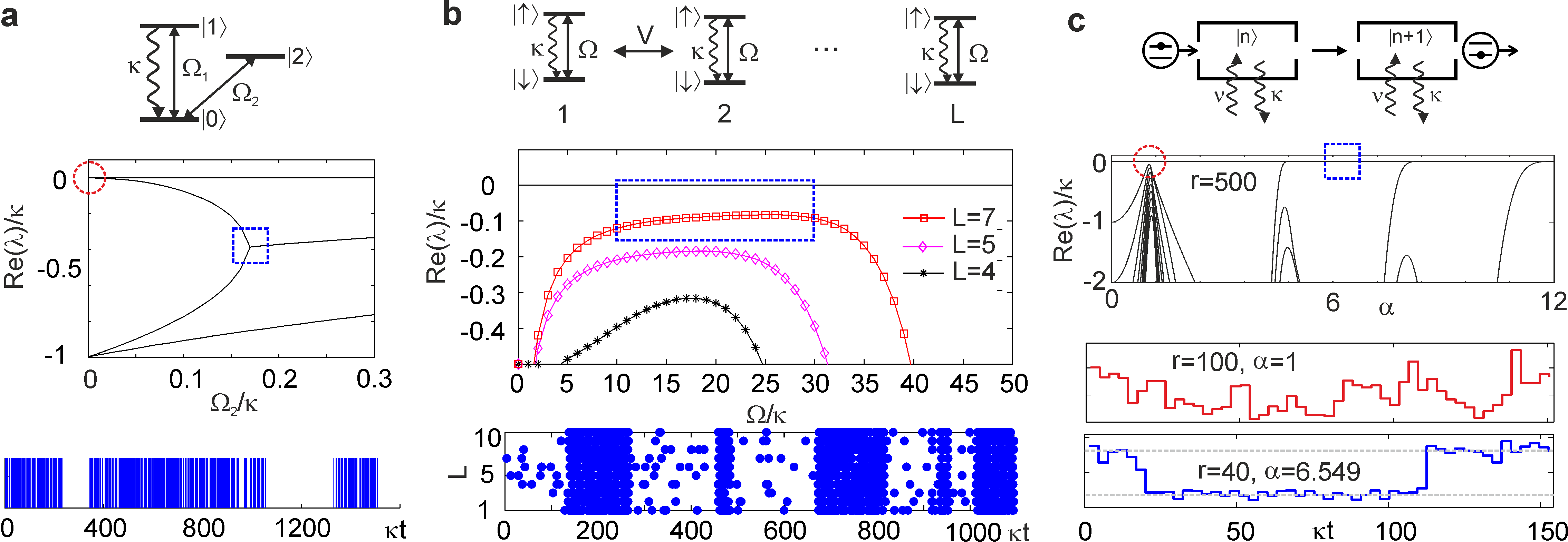}
\caption{Sketch of model systems, real part of the spectrum of the QMO, and sample trajectories (from top to bottom). \textbf{a:} The driven three-level system has a dynamical transition at the limiting value $\Omega_2=0$ (circled region). For $\Omega_2 \gtrsim 0$ quantum jumps are highly intermittent (we show the case $\Omega_2=0.01\kappa$ and $\Omega_1=4\kappa$). In the boxed region $\lambda_2$ acquires an imaginary part and time-correlations become oscillatory. \textbf{b:} The open Ising model with $L$ spins has a genuine many-body dynamical first-order transition and the spectral gap closes only in when $L\rightarrow\infty$. In the parameter regime chosen ($V=100\kappa$) the spectral gap becomes small for $10\kappa\leq\Omega\leq30\kappa$. In this near coexistence region of two dynamical phases one observes strongly intermittent behavior of quantum jumps (we show data for $L=10$, $\Omega=25\kappa$). \textbf{c:} The micromaser displays both first and second-order dynamical transitions. The two sample trajectories are taken at the second-order transition point (red circle/trajectory) and within the coexistence region of two dynamical phases (blue box/trajectory). In order to facilitate the representation we have collected quantum jumps in time bins of length $t_{\text{bin}} = 3/\kappa$.}
\label{fig:examples}
\end{figure*}

The limit of very long times is the ``thermodynamic limit'' of the associated spin problem. In this regime the two-time (connected) correlations of observables at positions $y$ and $y+x$ have the asymptotic form (see e.g. Ref. \cite{Schollwock05}),
$\left<\psi(f)\right| A^{(y)} B^{(y+x)}\left|\psi(f)\right>_c\propto \mathrm{Re} \mu_{2}^{x} = e^{-x/\xi} \cos(x \phi_{2}) $.
Here $\mu_2= |\mu_{2}|\exp(\pm i\phi_{2})$ denotes the eigenvalue(s)  with the second largest absolute value, of the transfer operator $E=\sum_{n=0}^N K^*_n \otimes K_n$  (or equivalently of $T_{\delta t}$), the largest eigenvalue being $\mu_{1}=1$ by the conservation of probability of the map (\ref{eq:map}). The correlation length $\xi$ is given by $\xi^{-1} = -\log |\mu_2|$. The correlations exhibit exponentially damped oscillations when the eigenvalue
$\mu_{2}$ is complex, in which case $\mu_{2}^{*}$ is also an eigenvalue of $T_{\delta t}$. It is more convenient to directly study the eigenvalues $\lambda$ of the QMO $\mathcal{W}$. In the limit $\delta t \rightarrow 0$ temporal correlations between quantum jumps behave as $\left<A(t)B(t+t')\right>_c\propto\exp(-t'/\tau) \cos{(\omega t')}$, with
$\tau \equiv -1/\mathrm{Re}(\lambda_{2})$ and $\omega \equiv \mathrm{Im}(\lambda_{2})$, where $\lambda_{2}$ is the eigenvalue of $\mathcal{W}$ with the second largest real part, Fig.\ \ref{fig:time_v_space}c (note, the eigenvalue with largest real part is $\lambda_{1}=0$).

Dynamical phase transitions will manifest in the closing of the spectral gap of the operator $\mathcal{W}$, i.e. $\lambda_{2}\rightarrow 0$ (see Fig. \ref{fig:time_v_space}c), a well-known feature in statistical mechanics which has also recently been used to characterize dynamical transitions in quantum spin models \cite{Kessler12} and the micromaser \cite{Garrahan11}. This closing of the gap can occur in qualitatively different ways giving rise to a different kinds of dynamical transitions. In the following we discuss this with four examples.

\textit{(i) Three-level system} - The three level system \cite{Plenio1998} as depicted in Fig.\ \ref{fig:examples}a is described by the Hamiltonian $H_\mathrm{3}=\Omega_1 \left|0\right>\!\left<1\right|+\Omega_2 \left|0\right>\!\left<2\right|+\mathrm{h.c.}$, and a single jump operator $L=\sqrt{\kappa}\left|0\right>\!\left<1\right|$. This system exhibits a dynamical phase transition at $\Omega_2=0$ (where $\lambda_2=0$),  which manifests itself in a strongly intermittent behavior of photon emission when $\Omega_2 \ll \Omega_{1}$ \cite{Marzoli&Cirac&Blatt&Zoller94}. The physical reason for this is that at $\Omega_2=0$ the system decouples into a driven two-level system undergoing frequent quantum jumps and an inactive single dark level. This corresponds to a twofold degeneracy of the leading eigenvalue (Fig.\ \ref{fig:examples}a), which is lifted when $\Omega_{2} > 0$. For $\Omega_2 \gtrsim 0$ the system can switch between these two phases on a timescale $\Omega^{-1}_2$, which results in strongly intermittent behavior reminiscent of a (smoothed) first-order transition \cite{Garrahan10}. From the perspective of the MPS this corresponds to the quantum phase transitions reported in Ref. \cite{Schon05}. Beyond the transition at $\Omega_2=0$, the three-level system features also a dynamical transition at finite $\Omega_2$ where time correlations become oscillatory (see Fig.\ \ref{fig:examples}a). Related transitions have been reported in an NMR experiment; see Ref.\ \cite{Alvarez06,Pastawski07}.

\textit{(ii) Dissipative Ising model} - In a many-body system the degeneracy leading to a phase transition does not need to be imposed externally, but can appear spontaneously in the thermodynamic limit. An example is the dissipative quantum Ising model \cite{Lee2012,Ates12} sketched in Fig.\ \ref{fig:examples}b. The Hamiltonian is that of a one-dimensional Ising chain in a transverse field: $H_\mathrm{I}=\Omega \sum^L_{k=1} \sigma^k_x+V \sum^L_{k=1} \sigma_z^k \sigma_z^{k+1}$ where $\sigma^k_\alpha$ are the Pauli spin matrices. The jump operators are $L_k=\sqrt{\kappa}\sigma^-_k$, which produce incoherent spin flips, $\left|\uparrow\right>\rightarrow\left|\downarrow\right>$, at a rate $\kappa$. In Fig. \ref{fig:examples}b we show the real part of the eigenvalue spectrum of the QMO $\mathcal{W}$ for various system sizes $L$. With increasing $L$ the spectral gap closes over an entire (coexistence) region in parameter space and we expect it to approach zero when $L\rightarrow\infty$ and hence the dimension of the transfer operator $E$ becomes infinite. This resembles a phase transition in the thermodynamical sense within the fictitious spin system. For finite $L$ and finite gap the system switches between two dynamical phases on long but finite timescales and quantum jump trajectories are strongly intermittent. Like in the three-level system, in this case the dynamical transition can be traced back to a bistable static behavior: When the emission of photons is plentiful the average magnetization is close to zero, while it is large and negative during the dark periods \cite{Ates12}.

\textit{(iii) Micromaser} - The micromaser features a critical point and a sequence of first order transitions \cite{Benson94,*Briegel&Englert&Sterpi&Walther,*Rempe&Walther90,*Englert02,Garrahan11}. It is modeled by a resonant single-mode cavity coupled to a finite temperature bath and pumped by excited 2-level atoms which are sent into the cavity with a constant rate $r$, Fig. \ref{fig:examples}c. The Hamiltonian is zero and there are four jump operators, two from the atom-cavity interaction, $L_1 = \sqrt{r}a^\dagger \frac{\sin \left( \phi \sqrt{aa^\dagger} \right)}{\sqrt{aa^\dagger}}$ and $L_2 = \sqrt{r} \cos \left( \phi \sqrt{aa^\dagger} \right)$, and two from the cavity-bath interaction, $L_3 = \sqrt{\kappa} a$ and $L_4 = \sqrt{\nu} a^\dagger$. Here $a,a^\dagger$ are the raising/lowering operators of the cavity mode, $\kappa$ and $\nu$ are the thermal relaxation and excitation rates, and $\phi$ encodes the atom-cavity interaction \cite{Garrahan11}. The spectrum of $\mathcal{W}$ is real and is depicted in Fig.\ \ref{fig:examples}c as a function of the ``pump'' parameter $\alpha=\phi/\sqrt{r}$. A sequence of first order transitions are visible beyond $\alpha=4$ and quantum jump trajectories (here we monitor quantum jumps associated to $L_1$) show the typical intermittent behavior. In contrast, in the vicinity of $\alpha=1$ the spectral gap closes in a way that makes the spectrum dense. This is the onset of a second-order phase transition which strictly only occurs in the limit $r\rightarrow\infty$ \cite{vanHorssen&Guta}. Typical quantum jump trajectories near the critical point fluctuate very strongly, as shown in Fig. \ref{fig:examples}c. Also the micromaser dynamical transitions are accompanied by a change in the statics, i.e. at the transition points the photon occupation of the cavity undergoes a sudden change: at first-order transitions the mean photon number switches between two largely distinct values, while at the critical point the variance of the photon number undergoes a jump \cite{Benson94,*Briegel&Englert&Sterpi&Walther,*Rempe&Walther90,*Englert02}. This seems to suggest that dynamical transitions can always be anticipated to occur by simply considering static or steady state properties. The next example proves that this is not the case.

\begin{figure}
\includegraphics[width=\columnwidth]{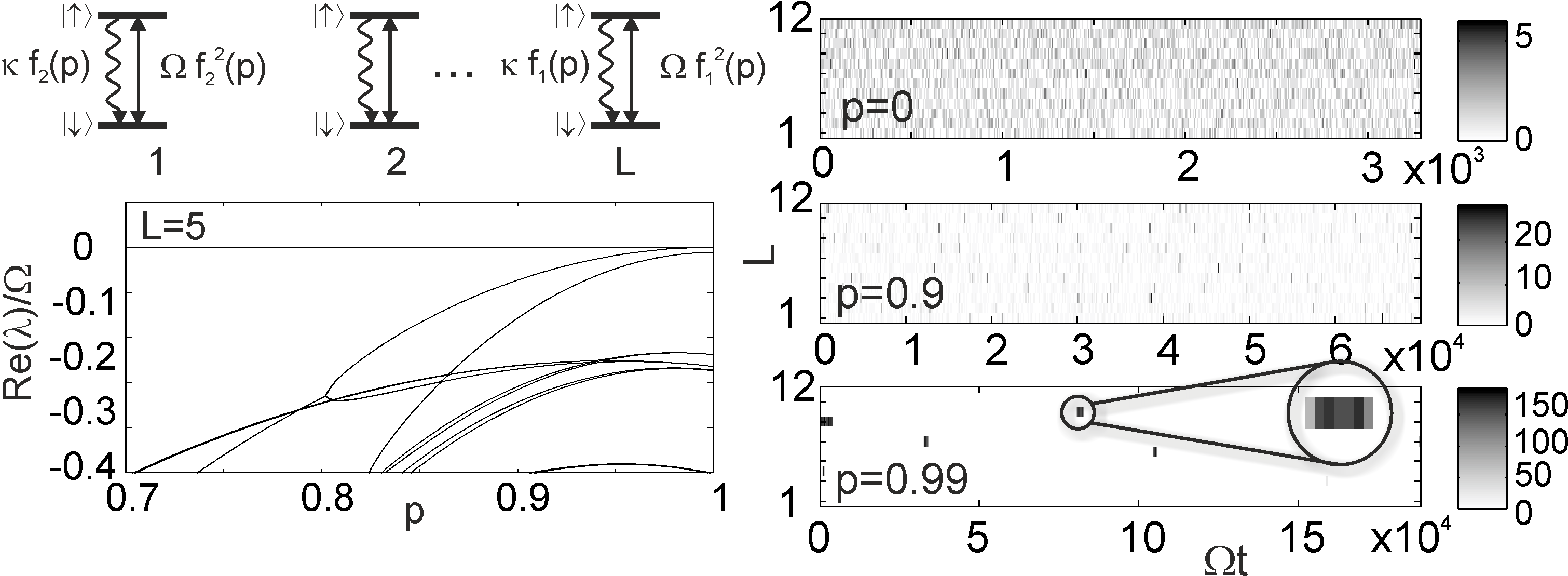}
\caption{Dissipative quantum glass model, with glassiness controlled by the parameter $p$: at $p=0$ system is non-interacting, for $p=1$ dynamics is fully constrained.  The spectral gap closes at $p=1$ (we show the case of $\kappa = 8\Omega$). Trajectories shown consist of $5000$ quantum jumps each. Dynamics changes drastically with $p$ but the stationary state remains invariant. Close to $p=1$ emission periods are localized in time and space (see magnified region). For ease of visibility jumps are collected in $500$ evenly spaced time bins.}
\label{fig:glass}
\end{figure}

\textit{(iv) Dissipative quantum glass} - This system is related to the dissipative quantum glass models of \cite{Olmos12}. It is a spin chain with Hamiltonian $H_\mathrm{g}=\Omega \sum_{k=1}^{L} \sigma^k_x f^2_{k+1}(p)$ and jump operators $L_k=\sqrt{\kappa}\sigma^-_k f_{k+1}(p)$ (see Fig. \ref{fig:glass}). The operators $f_k(p)$ are {\em kinetic constraints}.  If $f_k(p)=1$ the system is just a set of non-interacting two level systems with steady-state density matrix $\rho_\mathrm{ss}=\bigotimes_{k=1}^{L} \left[\left\{\frac{1}{2}+\frac{\omega\kappa}{\kappa^{2}+\omega^{2}}\right\} P_k + \left\{\frac{1}{2}-\frac{\omega\kappa}{\kappa^{2}+\omega^{2}}\right\} Q_k\right]$, where $\omega \equiv \sqrt{16 \Omega^{2}+\kappa^{2}}$, $Q_k=\frac{\mathbb{1}}{2} + \frac{\kappa}{2\omega}\sigma^k_z - \frac{2\Omega}{\omega} \sigma^k_y$, and $P_k=\mathbb{1}-Q_k$. Here the statics is clearly featureless. The problem becomes interacting and glassy if we choose the constraints to be $f_k(p)=p \, Q_k +(1-p)\mathbb{1}$ with $0\leq p\leq 1$: For $p=0$ we have the non-interacting problem and for $p=1$ a fully constrained quantum glass \cite{Olmos12}. When $p=1$ the state of the spin on site $k$ can only change if the state $\left|\phi\right>_{k+1}$ of its neighbor satisfies $Q_{k+1}\left|\phi\right>_{k+1}\neq 0$, leading to correlated dynamics in the system.  The parameter $p$ controls how glassy this dynamics is. Fig.\ \ref{fig:glass} shows that the spectral gap of the QMO would close at $p=1$. Here we expect a dynamical first-order phase transition to occur in analogy with classical constrained models \cite{Garrahan2007,Elmatad2007}. A crucial feature of this model is that the steady state for {\em any} value of $p$ is the trivial $\rho_\mathrm{ss}$, i.e., there is no change in the statics despite the change in the dynamics, as is evident from the example jump trajectories shown Fig.\ \ref{fig:glass}. As $p\rightarrow 1$ the system is most of the time inactive and quantum jumps become more and more localized in space and time, a phenomenon called dynamical heterogeneity which is a hallmark of glassy relaxation \cite{Ediger00}. In contrast to the three systems described above, here the dramatic dynamical change with $p$ is impossible to guess from static properties which are $p$ independent and trivial throughout.

Since the ensemble of trajectories is fully encoded in the MPS $|\Psi\rangle$ the sample trajectories of Fig.\ \ref{fig:glass} indicate a transition within the MPS as $p\rightarrow 1$. Usually such static quantum transitions are accompanied by a singularity, such as a logarithmic divergence in the entanglement entropy of large spin blocks \cite{Vidal&Lattore&Rico03,*Schuch&Wolf&Verstraete08}.
In our case this would be an entanglement between quantum jumps in subsequent long time segments. The entropy of a large block, however, is just two times the von Neumann entropy of the stationary state $S_{E}= -2{\rm Tr} (\rho_{ss} \log \rho_{ss})$ of the system, where $\rho_{ss} \equiv {\rm Tr}_{\psi} |\Psi\rangle \langle \Psi |$ \cite{Verstraete2005,*Wolf06}. Hence, due to the invariance of the stationary state with $p$, this entanglement measure does not detect the transition.

The reason for this is that in this problem---as is generic in both classical and quantum glasses, and likely to occur often in complex many-body systems---the appropriate fields driving the transition do not couple in an obvious way to static quantities, but do couple directly to time-integrated observables. An instance of these are the ``counting'' fields introduced when computing full counting statistics \cite{Levitov1996,*Bagrets2003,*Pilgram2003,*Flindt2008,*Esposito2009} of dynamical observables \cite{Garrahan2007,Garrahan10}.
Constrained models such as the one described here are known to exhibit dynamical transitions in trajectories, which are evident in the moment generating function (MGF) of the number of quantum jumps, $Z_{t}(s) \equiv \sum_{J} P_{t}(J) e^{-s J}$, where $P_{t}(J)$ is the probability of observing $J$ jumps in time $t$. In the $t \to \infty$ limit $Z_{t}(s)$ becomes singular at some value $s=s_{c}$ of the counting field \cite{Lecomte2007,Garrahan2007,Garrahan10}, indicating a phase transition in the ensemble of trajectories. At long times, the MGF is obtained from the largest eigenvalue of a deformation $\mathcal{W} \to\mathcal{W}_{s}$ of the QMO parametrized by $s$ \cite{Lecomte2007,Garrahan2007,Garrahan10}. The field $s$ therefore couples directly to the spectrum, so that a singularity of the MGF at $s_{c}$ indicates the existence of close to degenerate but distinct dynamical states. By driving $s$ one can single these states out; see Refs.\ \cite{Garrahan10,Garrahan11,Ates12} for details.

The MGF can be connected to the MPS through the norm, $Z_{t}(s) = \langle \Psi(s) |\Psi(s)\rangle$, of the deformation $|\Psi(s)\rangle \equiv e^{-s \hat{J}/2}|\Psi\rangle$, where $\hat{J}$ is an operator that counts the number of $n_{m>0}$ in a state $\left|n_1,..., n_M\right>$. In a thermodynamic analogy $Z_{t}(s)$ is like a partition sum over trajectories and $s$ a chemical potential which favors/disfavors quantum jumps.  The state $|\Psi(s)\rangle$ is thus a superposition of MPS like those of (\ref{eq:MPS}) but with each term weighed by a factor of $e^{{-s/2}}$ for each jump.  The eigenstate $\rho(s)$ of $\mathcal{W}_{s}$ corresponding to the largest eigenvalue is related to $|\Psi(s)\rangle$ through
${\rm Tr}_{\psi} |\Psi(s)\rangle \langle \Psi(s) | = Z_{t}(s) \rho(s)$, where ${\rm Tr}\rho(s)=1$ and $\rho(0) = \rho_{ss}$.  The entanglement entropy of this state, $\tilde{S}_{E} = -2{\rm Tr} [\rho(s) \log \rho(s)]$, will depend on $s$, and will display non-analytic behavior at $s_{c}$. At this level counting fields work in a similar manner to more standard static fields which drive phase transitions, but couple directly to the relevant dynamical order parameters that reveal transitions in quantum jump trajectories.  This perspective should be useful in the study of dynamical phase transitions in systems where they are not obviously connected to a change in spatial correlations.

\emph{Acknowledgments}.---
We acknowledge discussions with B. Olmos. The work was supported by EPSRC Grant no.\ EP/J009776/1 and Fellowship EP/E052290/1.

\bibliographystyle{apsrev4-1}

\end{document}